\documentclass[a4paper,11pt]{article}
\usepackage{pos}
\usepackage{multirow}
\usepackage{xcolor}

\title{Meeting the Challenges for Relic Neutrino Detection}

\author[a]{P. S. Bhupal Dev}
\author*[b]{Amarjit Soni}

\affiliation[a]{Department of Physics and McDonnell Center for the Space Sciences, Washington University, St.\,Louis, MO 63130, USA}

\affiliation[b]{Physics Department, Brookhaven National Laboratory, Upton, NY 11973, USA}

\emailAdd{bdev@wustl.edu}
\emailAdd{adlersoni@gmail.com}

\abstract{Inspired by Gounaris-Sakurai and Lee-Zumino, we postulate that the weak vector and axial vector currents are dominated by $J^{PC} = 1^{--}$ and $1^{++}$ resonances respectively in the appropriate channels of $\nu + \bar \nu$ annihilation into quark-antiquark pairs when an ultrahigh-energy incoming $\nu \ (\bar \nu)$ strikes a relic $\bar \nu \ (\nu)$. Despite this and some other ideas, it appears the detection of relic neutrinos with just the Standard Model interactions seems extremely difficult at existing or future neutrino telescopes. Thus any positive signal would be due to some non-standard interactions of neutrinos.}

\FullConference{%
 The European Physical Society Conference on High Energy Physics (EPS-HEP2021),\\
  26-30 July 2021\\
  Online conference, jointly organized by Universität Hamburg and the research center DESY 
}

\begin{document}
\maketitle

\section{Introduction}
Clearly the detection of cosmic (relic) neutrino background is an extremely important problem in Particle Physics, and as is well known, it is extremely challenging and that is what makes it very interesting~\cite{Gelmini:2004hg, 10.3389/fphy.2014.00030, Vogel:2015vfa}. In this proceeding, we will study the possibility of detecting relic neutrinos via their resonant annihilation with ultrahigh energy (UHE) neutrinos. Recall that the effective temperature of the cosmic neutrino background is $T_\nu\simeq 1.945~{\rm K}\simeq 1.7 \times 10^{-4}~{\rm eV}$. From the neutrino oscillation data, we of course only know two mass-square differences, while the absolute value of the neutrino mass is still unknown. For our numerical illustrations below, we will assume that the neutrino mass is 
$\approx \sqrt {\Delta m_{\rm atm}^2} \simeq 0.05~{\rm eV}$ which is clearly much larger than the  effective temperature for relic neutrino, so it is almost sitting at rest. But in reality there should be three masses for the three neutrino species. As a result we should expect three curves 
(or peaks/dips, if relevant) as functions of incident colliding neutrino energies.

Because the neutrino masses are of ${\cal O}(\lesssim 0.05~{\rm eV})$ the colliding neutrino has to possess ultra-high energies even for production 
of the lightest Standard Model (SM) resonances, such as $e^{+}e^{-}$ or $\mu^{+}\mu^{-}$ pairs, or $\rho^0,\ \phi$ mesons etc; see Table~\ref{table:1}. Thus, the squared center-of-mass energy needed is 
\begin{eqnarray}
s = 2 m_{\nu} E_{\nu} \, , \quad {\rm or}, \quad 
E_{\nu} \simeq \left(\frac{s}{{\rm GeV}^2}\right)  \times 10^{10}~{\rm GeV} \, ,
\end{eqnarray}
where $E_{\nu}$ is the incoming UHE neutrino energy.
Because such large energies are needed the incident flux rates tend to be low. This is the main reason for exceedingly small event rates, as shown below. 
\begin{table}[h]
\centering
\begin{tabular}{c|c|c}
	\hline
	\hline
	Particle produced & Incident neutrino energy needed & comments \\
	\hline
	$e^+e^{-}$ & 10 TeV & \multirow{3}{*}{IceCube detected up to a few PeV}  \\
	$\mu^{+} \mu^{-} $& 447 PeV &    \\
	$\pi^{+} \pi^{-}$ & 784 PeV &   \\ \hline
	$\rho$ & 6 EeV  & \multirow{5}{*}{Balloon experiments detected up to about EeV }  \\
	$\phi$ & 10 EeV &      \\
	$J/\Psi$  & 96 EeV  &      \\
	$\Upsilon$ & 1 ZeV &      \\
	$ Z$  &   83 ZeV  &       \\ \hline
	NSI $Z^\prime$ (50 MeV)  & 25 PeV &  \multirow{2}{*}{Detectable at future neutrino experiments} \\
 	NSI $Z^{\prime}$ (1.8 GeV) & 32 EeV &   \\ \hline\hline
\end{tabular}
\caption{Incident neutrino energies needed for resonant production of various SM particles, as well as two examples of beyond SM resonances. Recall that 1 EeV = $10^{18}$ eV and 1 ZeV=$10^{21}$ eV.}
\label{table:1}
\end{table}

\section{Standard Model Resonances}
 Recall that in the  SM~\cite{ParticleDataGroup:2020ssz}, left-handed components of $\nu$ and $\ell^{-}$ are doublets of $SU(2)_L$: $\psi_i=(\nu_i, \ \ell_i^-)^T$, where $i=1,2,3$ is the family index. After electroweak symmetry breaking, the relevant piece of the Lagrangian reads 
\begin{align}
{\cal L} \supset -e Q_i \bar \psi_i \gamma^\mu \psi_i A_\mu - \frac{g}{2 \cos\theta_w} \bar \psi_i \gamma^\mu (g_V^i  - g_A^i \gamma_5 ) \psi_i Z_\mu \, , 
\end{align}
where $A_\mu$ and $Z_\mu$ are the photon and the $Z$-boson fields respectively, $g$ is the $SU(2)_L$ gauge coupling, $\theta_w$ is the weak mixing angle,  $e=g\sin\theta_w$ is the positron electric charge, $Q_i$ is the fermion charge relative to the positron, and $g_{V,A}$ are the vector and axial-vector couplings for the $Z$-boson.

We are all very familiar with $e^{+}e^{-}$ collisions. For the center-of-mass energy $\sqrt{s}$ much less than the $Z$ mass, 
the $e^+e^-\to f\bar{f}$ ($f$ being any SM fermion) cross-section falls as $1/s$ for dimensional reasons. Therefore, the ratio $R$ of the cross-section into quark-antiquark pairs divided by the cross-section into $\mu^{+}\mu^{-}$ pairs is a constant that depends on the underlying quark degrees of freedom; see e.g.~Figure 52.2 in PDG~\cite{ParticleDataGroup:2020ssz}.  One can  readily see that for $s\ll m_Z^2$, the $e^{+}e^{-}$ cross-section is dominated by the vector meson resonances with $J^{PC}=1^{--}$, such as $\rho$, $\omega$, $\phi$, $J/\Psi$,  $\Upsilon$ etc,  as predicted long ago by Gounaris and Sakurai ~\cite{Gounaris:1968mw}. 


In much the same way, when it comes  to $\nu \bar \nu$ collisions, the weak current now is either of vector or axial-vector type. There is nothing particularly sacred about the vectors only. Specifically, inspired by the general idea of Lee and Zumino~\cite{Lee:1967iv} we postulate that in analogy with the weak vector current, the axial-vector current will likewise be dominated by the corresponding $J^{PC}=1^{++}$ resonances, such as $a_1(1260)$, $f_1(1285)$, $f_1(1420)$, $f_1(1510)$, $\chi_{c1} (3872)$,  $\chi_{b1}$, etc. 

\begin{figure}[t!]
\centering
\includegraphics[width=0.49\textwidth]{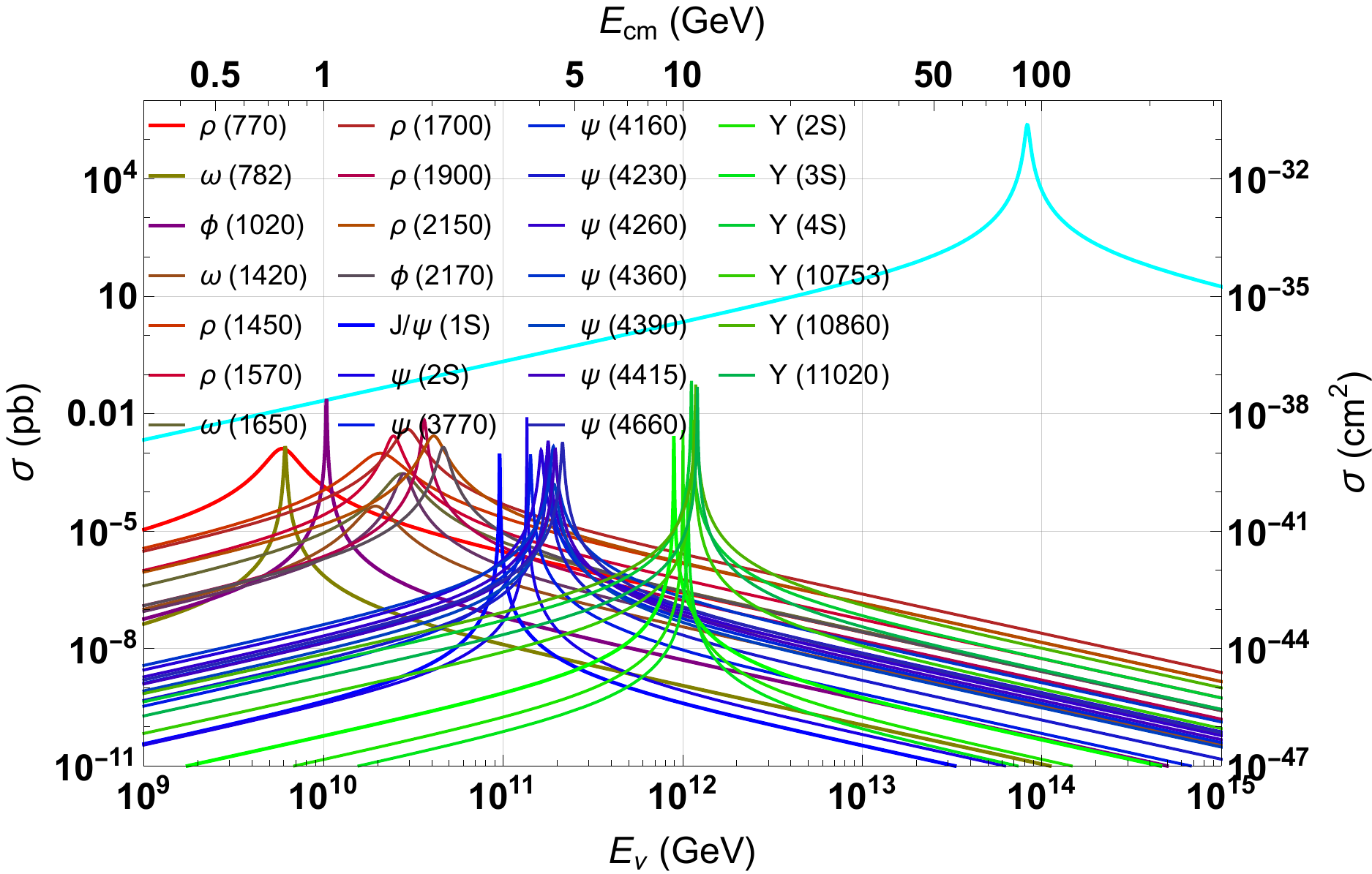}
\includegraphics[width=0.49\textwidth]{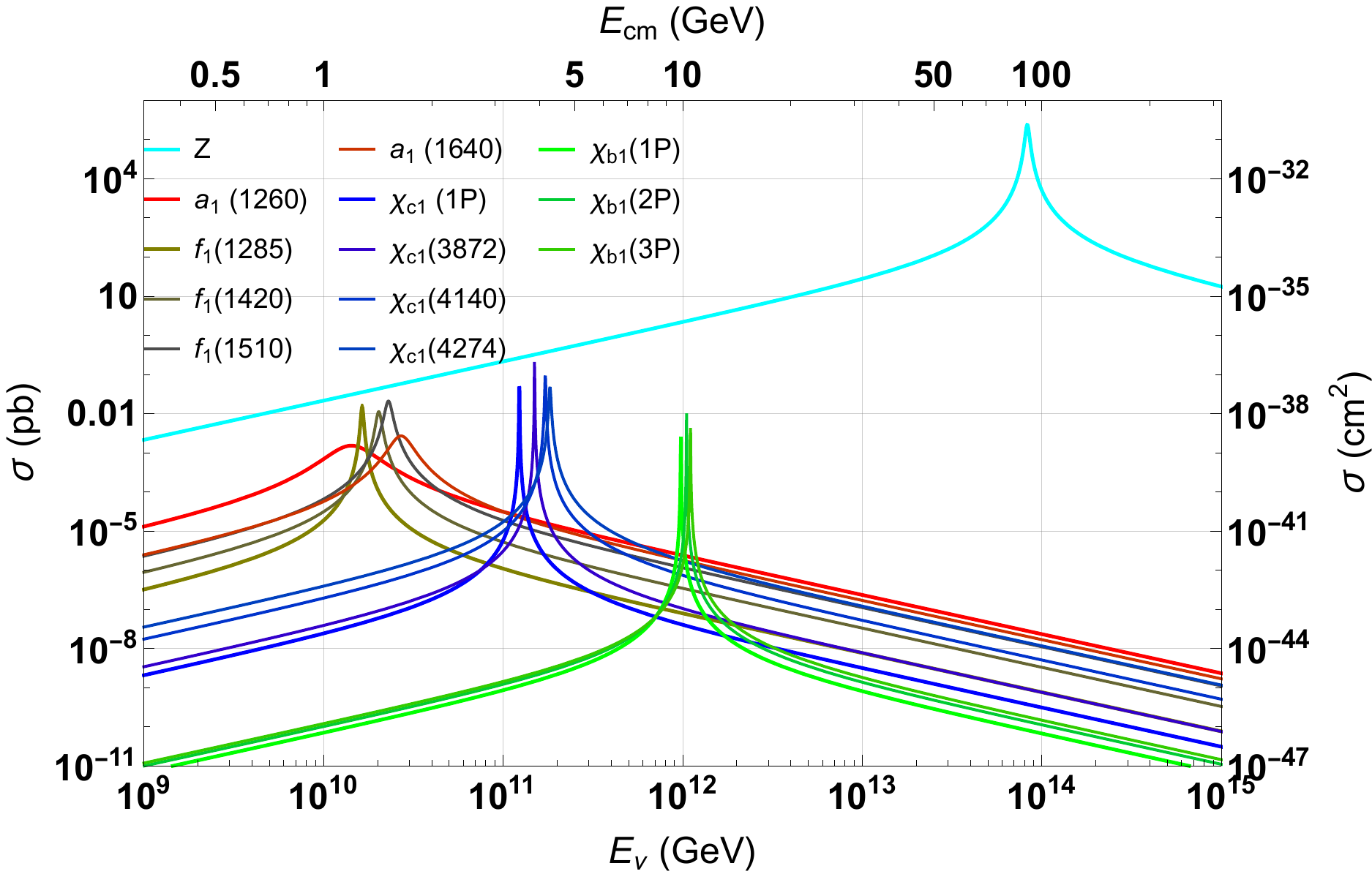}
\caption{Vector (left) and axial-vector (right) meson resonances in $\nu-\bar{\nu}$ scattering. The $Z$-resonance is also shown (in cyan) for comparison.} \label{fig:cross}
\end{figure}

We use the Breit-Wigner resonance formula~\cite{ParticleDataGroup:2020ssz} for calculating the cross-section:
\begin{align}
    \sigma(\nu_i\bar{\nu}_i\to X^*\to {\rm anything}) = \frac{8\pi}{m_X^2}{\rm BR}(X\to \nu_i\bar\nu_i){\rm BR}(X\to {\rm anything})\frac{s\Gamma_X^2}{\left(s-m_X^2\right)^2+\frac{s^2\Gamma_X^2}{m_X^2}} \, ,
    \label{eq:crossSM}
\end{align}
where BR stands for the branching ratio of the resonance particle $X$ to neutrinos and other final states. 
For the vector meson case, we use the following expressions for ${\rm BR}(V\to \nu_i\bar\nu_i)$:
\begin{align}
   \Gamma(\rho\to \nu_i\bar\nu_i) & =  \frac{G_F^2}{24\pi}\left(1-2\sin^2\theta_w\right)^2f_\rho^2m_\rho^3 \, , \\
   \Gamma(\omega\to \nu_i\bar\nu_i) & = \frac{G_F^2}{96\pi}\left(-\frac{4}{3}\sin^2\theta_w\right)^2 f_\omega^2 m_\omega^3 \, , \\
  \Gamma(\phi\to \nu_i\bar\nu_i) & = \frac{G_F^2}{96\pi}\left(-1+\frac{4}{3}\sin^2\theta_w\right)^2 f_\phi^2 m_\phi^3 \, , \\
   \Gamma(\Psi\to \nu_i\bar\nu_i) & = \frac{G_F^2}{96\pi}\left(1-\frac{8}{3}\sin^2\theta_w\right)^2 f_\Psi^2 m_\Psi^3 \, , \\
    \Gamma(\Upsilon\to \nu_i\bar\nu_i) & = \frac{G_F^2}{96\pi}\left(-1+\frac{4}{3}\sin^2\theta_w\right)^2 f_\Upsilon^2 m_\Upsilon^3 \, , 
\end{align}
where $G_F$ is the Fermi constant and $f_X$ is the $X$-meson decay constant. The meson masses, decay widths and decay constants (when directly available) are taken from the PDG~\cite{ParticleDataGroup:2020ssz}. For the higher resonances, we have estimated the decay constants using the ratio of their $V\to e^+e^-$ decay rates from the PDG and using the formula
\begin{align}
    \Gamma(V\to e^+e^-)  =  \frac{4\pi}{3}\frac{\alpha^2}{m_V}f_V^2c_V \, ,
\end{align}
where the coefficients $c_V$ are given in Appendix C of Ref.~\cite{Bharucha:2015bzk}. For instance, the decay constant for $\omega' \ (1420)$ is estimated as
\begin{align}
    \left(\frac{f_{\omega'}}{f_\omega}\right)^2 = \frac{\Gamma(\omega'\to e^+e^-)}{\Gamma(\omega\to e^+e^-)}\frac{m_{\omega'}}{m_\omega} = \frac{{\rm BR}(\omega'\to e^+e^-)}{{\rm BR}(\omega\to e^+e^-)} \frac{\Gamma_{\omega'}}{\Gamma_\omega}\frac{m_{\omega'}}{m_\omega} = \frac{6.6\times 10^{-7}}{7.39\times 10^{-5}}\frac{290~{\rm MeV}}{8.68~{\rm MeV}}\frac{1410~{\rm MeV}}{782.66~{\rm MeV}}.\nonumber
\end{align}
The cross-sections Eq.~\ref{eq:crossSM} as a function of the incoming neutrino energy (or the center-of-energy) are shown in Figure~\ref{fig:cross} for the SM vector (left panel) and axial-vector (right panel) meson resonances. The $Z$-peak is also shown for comparison. It is clear that none of these meson resonances go beyond the $Z$-continuum. This makes the observational prospects of any of these resonances in neutrino scattering rather bleak, as we will show in the next section (see also Ref.~\cite{Paschos:2002sj}).

\section{Event Spectrum} 
\begin{figure}[t!]
\centering
\includegraphics[width=0.49\textwidth]{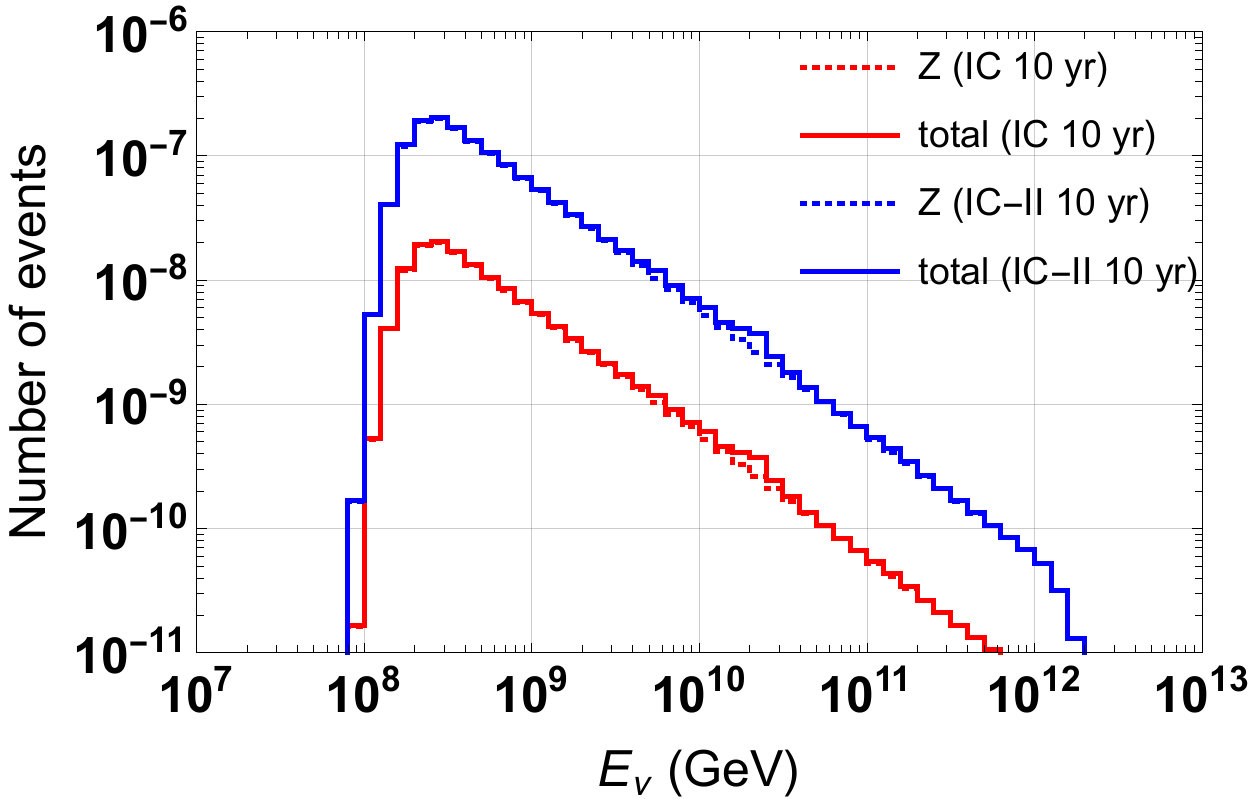}
\includegraphics[width=0.49\textwidth]{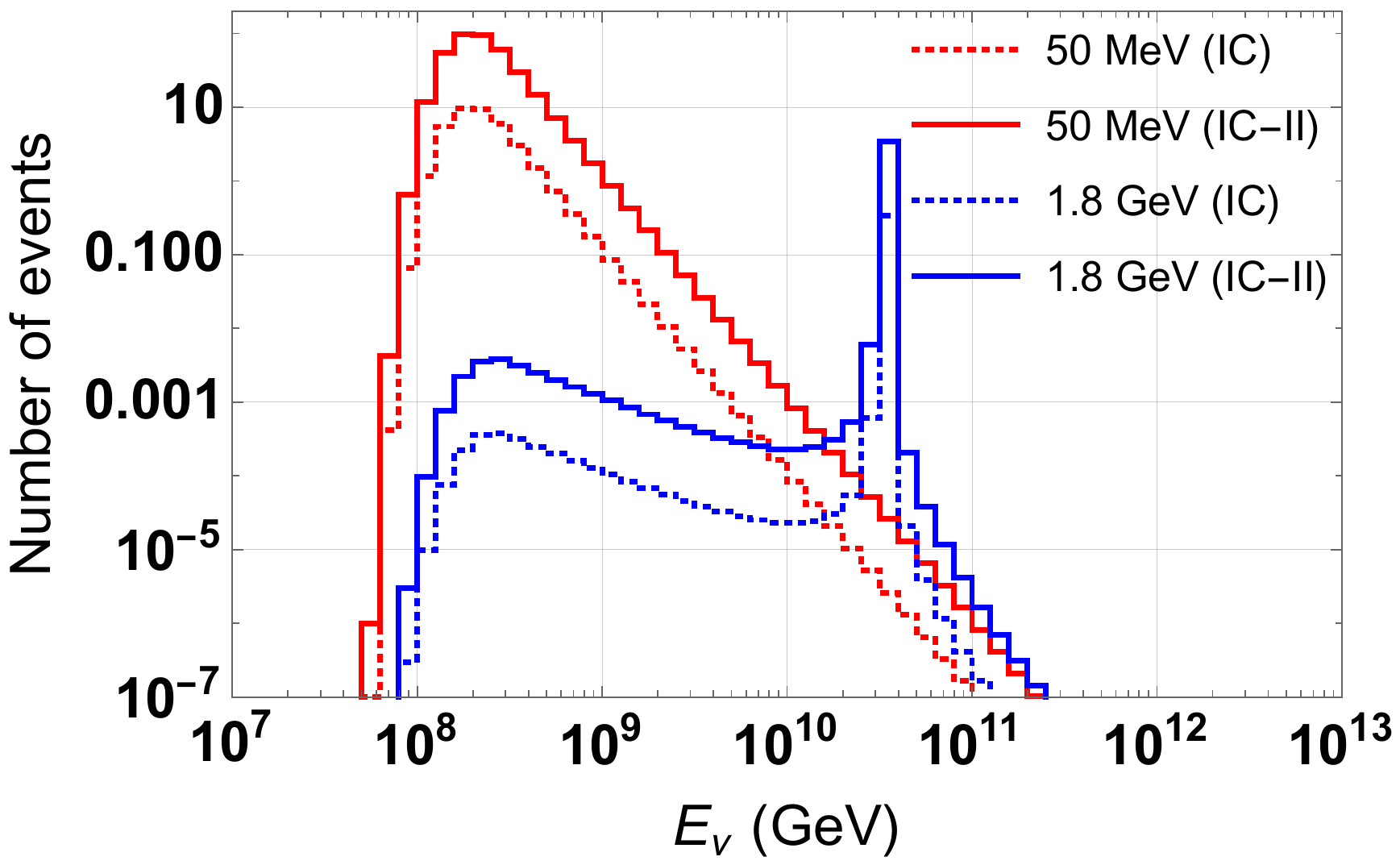}
\caption{Event rate at current and future versions of IceCube. Left panel: SM only. Right panel: With NSI.} 
\label{fig:event}
\end{figure}

Given the cross section and the incident neutrino flux, the total number of events in a neutrino telescope like IceCube can be calculated as 
\begin{align}
    N=T\cdot N_A \cdot \Omega \cdot V\cdot \int_{E_{\rm min}}^{E_{\rm max}}dE \Phi(E)\sigma(E) \, ,
    \label{eq:event}
\end{align}
where $T$ is the exposure time (which we take 10 years), $V=1~{\rm km}^3$ is the detector volume, $N_A=6.022\times 10^{23}~{\rm cm}^{-3}$ is the water-equivalent of Avogadro number,  $\Omega=2\pi$ sr is the solid angle of coverage (only events above horizon are considered; the UHE neutrinos coming from below will be severely attenuated by Earth), $\Phi$ is the cosmogenic neutrino flux (in units of ${\rm GeV}^{-1}{\rm cm}^{-2}{\rm s}^{-1}{\rm sr}^{-1}$) for which we use the model given in Ref.~\cite{Ahlers:2010fw}. The ranges of integration $E_{\rm min}$ and $E_{\rm max}$ specify the bin size in the event spectrum. 
The event spectrum for the sum of all SM resonances is shown in Figure~\ref{fig:event} left panel. The dashed lines are for the $Z$-resonance alone and the solid ones are including all the vector and axial-vector resonances shown in Figure~\ref{fig:cross}. The red lines are for the current IceCube detector volume and the blue ones are for a future IceCube Gen-II with ten times the IceCube volume.  It is clear from this figure that with the detectors of today or even with much large ones that may be available in the foreseeable future, it is extremely  unlikely that relic neutrinos will be detected with the SM interactions only. Thus, if such detectors are able to "see" relic neutrinos they are very likely to have some non-standard interactions (NSI). We will study two examples of NSI in the next section to serve as illustrations.   


\section{Two examples of NSI}
We consider a simple NSI model where a light vector boson $Z'$ couples to the neutrinos: 
\begin{align}
    {\cal L} \supset g_L'\bar\nu_i\gamma^\mu P_L \nu_j Z'_\mu \, .
\end{align}
Note that the couplings can be either flavor-diagonal ($i=j$) or non-diagonal ($i\neq j$). The corresponding $\nu\bar\nu$ cross section is given by~\cite{Altmannshofer:2016brv} 
\begin{align}
    \sigma(s)=\frac{g_L'^4}{3\pi}\frac{s}{\left(s-m_{Z'}^2\right)^2+m_{Z'}^2\Gamma_{Z'}^2} \, ,
\end{align}
where $\Gamma_{Z'}=g_L'^2m_{Z'}/24\pi$ is the $Z'$ decay width (including the charged lepton channels). The corresponding event spectrum, calculated using Eq.~\eqref{eq:event}, is shown in Figure~\ref{fig:event} right panel for a benchmark value of $g_L'=0.01$ and for two different values of $m_{Z'}=50$ MeV (red) and 1.8 GeV (blue). As in the left panel, the dashed (solid) lines are for 10 years of IceCube (IceCube Gen-II) exposure. We see that the number of events in the $\nu\bar{\nu}$ scattering can be sizable in presence of this NSI, but it remains to be seen whether such large NSI effects are robustly excluded by other laboratory experiments. This is currently under investigation.

\section{ Summary}
\begin{itemize}
    \item Detection of relic neutrinos is clearly extremely important and also extremely challenging.
    \item We have studied the prospects of relic neutrino detection via their resonant scattering with UHE neutrinos. 
    \item Since neutrino masses are so very small, the colliding neutrino energies required even for production of muons, pions, $\rho$ etc are hundreds of PeVs, consequently the incoming cosmogenic fluxes are very low.
    \item Inspired by the works of Gounaris-Sakurai~\cite{Gounaris:1968mw},
    and bearing in mind the work of Lee-Zumino (see, ``Field-current identities"~\cite{Lee:1967iv}) in regard to photons and QED, we generalize it to the weak vector and
    weak axial-vector currents. Despite these additions, the cross-section still appears to be dominated by the Z-boson. However, the collision energy needed for the $Z$ is extremely high, $\approx 10^{14} ~{\rm GeV}$, as discussed originally in Ref.~\cite{Weiler:1982qy}  and the corresponding flux exceedingly low rendering the experimental observation a daunting challenge.
    
    \item Consequently, any significant signal of relic neutrinos that is seen in a detector, for the foreseeable future, is very likely to be a result of NSI. Since there now exist  serious anomalies in muon $(g-2)$, in flavor physics and in neutrino physics~\cite{Fischer:2021sqw}, NSI may well be affecting neutrinos~\cite{Proceedings:2019qno}. Motivated by such thoughts, we gave two examples of light $Z'$ of mass 50 MeV and of 1.8 GeV coupling to neutrinos, in which case it turns out that the $\nu\bar{\nu}$ scattering rates can in principle be enhanced to an observable level. 
    
\end{itemize}

\section*{Acknowledgments}
This work of BD was supported in part by the U.S. Department of Energy under Grant No. DE-SC0017987 and by a Fermilab Intensity Frontier Fellowship. The work of AS was supported in part by the U.S. DOE contract \#DE-SC0012704.

\bibliographystyle{apsrev4-1}
\bibliography{references}  

\end{document}